%% file: Manuscript.tex
\documentclass[comsoc, journal]{IEEEtran}
\IEEEoverridecommandlockouts
\pdfoutput=1
\usepackage[numbers]{natbib}
\setcitestyle{open={[},close={]}}

\usepackage[cmintegrals]{newtxmath}
\usepackage{amsmath}

\usepackage{graphicx}
\usepackage{textcomp}
\usepackage{xcolor}
\usepackage{float}
\usepackage{array}
\usepackage[font=footnotesize,labelfont=bf]{caption}
\usepackage{optidef}
\usepackage{subcaption}
\usepackage[utf8]{inputenc}
\usepackage{algorithm}
\usepackage{algpseudocode}
\usepackage{color, xcolor, colortbl}
\newcommand{\revision}{\color{black}}

\renewcommand\IEEEkeywordsname{Keywords}

\def\BibTeX{{\rm B\kern-.05em{\sc i\kern-.025em b}\kern-.08em
    T\kern-.1667em\lower.7ex\hbox{E}\kern-.125emX}}
\begin{document}
\bstctlcite{IEEEexample:BSTcontrol}

\title{Clustered Vehicular Federated Learning: Process and Optimization}

\author{
	\IEEEauthorblockN{
				Afaf Taïk \IEEEmembership{Student Member, IEEE}, Zoubeir Mlika \IEEEmembership{Member, IEEE}  and
		Soumaya Cherkaoui \IEEEmembership{Senior Member, IEEE} 
	}
	
	\thanks{Authors are with the INTERLAB Research Laboratory, Faculty of Engineering, Department of Electrical Engineering and Computer Engineering, Université de Sherbrooke, Sherbrooke (QC) J1K 2R1, Canada.}
}

\markboth{}{} 

\maketitle

\begin{abstract}
Federated Learning (FL) is expected to play a prominent role for privacy-preserving machine learning (ML) in autonomous vehicles.  FL involves the collaborative training of a single ML model among edge devices on their distributed datasets while keeping data locally. 
While FL requires less communication compared to classical distributed learning, it remains hard to scale for large models. In vehicular networks, FL must be adapted to the limited communication resources, the mobility of the edge nodes, and the statistical heterogeneity of data distributions. Indeed, a judicious utilization of the communication resources alongside new perceptive learning-oriented methods are vital. 
To this end, we propose a new architecture for vehicular FL and  corresponding learning and scheduling processes. The architecture utilizes vehicular-to-vehicular(V2V) resources to bypass the communication bottleneck where clusters of vehicles train models simultaneously and only the aggregate of each cluster is sent to the multi-access edge (MEC) server. The cluster formation is adapted for single and multi-task learning, and takes into account both communication and learning aspects. We show through simulations that the proposed process is capable of improving the learning accuracy in several non-independent and-identically-distributed (non-i.i.d) and unbalanced datasets distributions, under mobility constraints, in comparison to standard FL. 

\end{abstract}


\IEEEpeerreviewmaketitle

\input{1_introduction}

\input{2_background}

\input{3_model}

\input{4_design}

\input{5_evaluation}

\input{6_conclusion}

\section*{Acknowledgments}
The authors would like to thank the Natural Sciences and Engineering Research Council of Canada, for the financial support of this research.

\bibliographystyle{IEEEtran}
\bibliography{references}

\end{document}

%% file: 1_introduction.tex
\section{Introduction}
Autonomous driving (AD) requires little-to-no human interactions to build an intelligent transportation system (ITS). Consequently, AD helps in reducing accidents caused by human driving errors. Artificial intelligence (AI) plays an essential role in AD by empowering several applications such as object detection and tracking through machine learning (ML) techniques \cite{yurtsever_survey_2020, grigorescu_survey_2020}.

With the raise of AI research and deployment over the last decade, the development of autonomous vehicles has seen significant advancements. Indeed, vehicle manufacturers put a lot of effort to deploy AI schemes aiming to achieve human-level situational awareness. However, owing to technical difficulties and several ethical and legal challenges, it is still challenging for vehicles to achieve full autonomy. In fact, autonomous vehicles need to fulfill strict requirements of reliability and efficiency, and achieve high levels of situational awareness. Vehicle manufacturers are deploying efforts to achieve these goals. Autonomous vehicles will be capable of sensing their network environment using embedded sensors and share information with other vehicles and equipment through wireless communication. Autonomous vehicles can be equipped with LiDAR sensors, camera sensors, and radar sensors that collect important amounts of data to share with the vehicular network. 

With the prevalence of connected vehicles and the transition toward autonomy, it is expected that vehicles will no longer rely only on locally collected data for localization and operation. Instead, enhanced situational awareness can be attained through exchanging raw and processed sensor data among large networks of interconnected vehicles \cite{lu_blockchain_2020}. In contrast to status data sharing, sensor data sharing becomes a pivotal operation for different safety applications, such as HD map building \cite{szabo_smartphone_2019} and extended perception \cite{kim_cooperative_2016}.  These data are also necessary to produce or enhance ML models that will be capable of performing AD tasks, such as dynamically adjusting the vehicle's speed, braking, and steering, by observing their surrounding environment.

Nonetheless, extensive sensor data sharing raises alarming privacy issues since vehicle sensor sharing involves sharing raw and processed data among vehicles. These data expose sensitive information about the vehicle, the driver, and the passengers, and could be used in a harmful way by a malicious entity. While privacy in vehicle status sharing has been already been extensively addressed and regulated by vehicle manufacturers---through a dynamic change of media access control (MAC) address and data anonymization, these regulations have not been extended to sensor data sharing. Moreover, to attain fully AD and enhance the overall ML models' performance, the deployed ML/AI models in the vehicle need to be updated and improved periodically by original equipment manufacturers (OEM). This requires the vehicles to upload the collected data to the OEMs, which further violates data privacy. Indeed, when data is uploaded to multi-access edge computing (MEC) \cite{filali_multi-access_2020,amine_mec} servers, or to the cloud, it may be subject to be malicious interception and misuse.

Federated learning (FL)\cite{r8} has emerged as an attractive solution for privacy-preserving ML. FL consists of the collaborative training of ML models among edge devices without data-sharing, which makes it a promising solution for the continuous improvement of ML models in AD. Indeed, with FL, edge devices share their models parameters instead of their private data and then the models are aggregated at MEC servers to obtain a global accurate model. 

When FL is used in a vehicular network context, a centralized entity (e.g., a MEC server) initializes a model and distributes it among participant vehicles. Each vehicle then trains the model using local data and sends the resulting model parameters to the central entity for aggregation. 

The predominant FL training scheme is a synchronous aggregation. Accordingly, the MEC server waits for all vehicles to send their updates before aggregating them. 

The assumption of FL is that the goal for participating end devices (also called end users throughout the article)  is to approximate the same global function. Nevertheless, this is not the case for non-i.i.d data, particularly in the case of competing objectives, where a single joint model cannot be optimal for all end devices simultaneously. Consequently, clustering \cite{sattler_clustered_2020,briggs_federated_2020} was proposed to group users with similar objectives and build multiple versions of the trained model. However, these works suppose the availability of all the end users and require their participation in the training for cluster-formation. Therefore, even if vehicle clustering for FL is interesting for the above mentioned reasons,  due to the high-speed mobility, Doppler effect, and frequent handover (short inter-connection times), not all vehicle updates can be collected at the MEC servers. Further, due to the different mobility patterns, not all vehicles can have strong signal quality with the MEC servers. As a result, participating vehicles should be carefully selected and communication must be efficiently scheduled. 

Vehicle-to-vehicle (V2V) communication offers a new opportunity for FL deployment that bypasses the communication bottleneck with the MEC server\cite{meysam3}. A cluster of vehicles can collaboratively train models and a chosen cluster-head can aggregate their updates so as only one model is sent to the MEC server. To achieve this, two main questions need to be addressed: how to adequately form FL clusters under mobility constraints; and how to select the cluster-heads in such settings. 

In this article, we propose a cluster-based scheme for FL in vehicular networks. The clustering scheme consists of grouping vehicles with common characteristics, not only in terms of direction and velocity, but also from a learning perspective through the evaluation of the updates' similarity. Thus, the proposed scheme allows to accelerate the models' training through ensuring (i) a larger number of  participants (ii) possibility to train several models to adapt to non-i.i.d and unbalanced data distributions. 

The main contributions of this article can be summarized as follows:
\begin{itemize}
    \item[1)] we design an architecture and corresponding FL process for clustered FL in vehicular environments;
    \item[2)] we formulate a joint cluster-head selection and resource block allocation problem taking into account mobility and data properties;
    \item[3)] we formulate a matching problem for cluster formation taking into account mobility and model preferences;    
    \item[4)] we prove that the cluster-head problem is NP-hard and we propose a greedy algorithm to solve it;
    \item[6)] we evaluate the proposed scheme through extensive simulations. 
\end{itemize}

The remainder of this article is organized as follows: 
In Section II, we present the background for FL and related work. In Section III, we present the design of the learning process and considered system model components.  In Section IV, we formulate the cluster-head selection and vehicle association problems, and we present the proposed solution. Simulation results are presented in Section V. At last, conclusions and future work are presented in Section VI.

%% file: 2_background.tex
\section{Background}
\begin{table}[t]
	\centering
	\caption{List of Notations.}
	
\begin{tabular}[h]{l|p{4.5cm}}
	\hline
	\textbf{Notations}&\textbf{Description}\\
	\hline
	$T_k$& Rate of stay of vehicle $\it{k}$\\
	$Total_{RB}$& Total available resource blocks\\
	$s$& Model size\\
	$\epsilon$& Number or local epochs\\
	$g$ & Global model\\	
	$\theta_k$ & Model update of vehicle $\it{k}$\\
	$t_{k}^{train}$ & Training time for vehicle $\it{k}$\\
	$r_k$ & Achievable data rate of vehicle $k$\\
	$t_{k}^{up}$ & Upload time for vehicle $\it{k}$\\	
	$P_k$ & Transmit power of vehicle $k$\\
	$N_0$& Power spectral density of the Gaussian noise\\
	$\left | D_k \right |$ & vehicle $k$'s dataset size\\
	$I_{k}$& Data-diversity of vehicle $\it{k}$\\
	$R_{k,h}$& Relationship of vehicles $\it{k}$ and $\it{h}$\\	
	\hline
\end{tabular}
	\label{tab:notation}
\end{table} 

In this section, we first present a background on FL and challenges tackled in this paper, then we present related work that enables and motivates our work.
\subsection{Federated Learning}
FL is a privacy-preserving distributed training framework, which consists of the collaborative training of a single ML model among different participants (e.g.,IoT devices) on their local datasets. 
The training is an iterative process that starts with the global model initialization by a centralized entity (e.g., a server). In every communication round $\it{i}$ , a selected subset of $\it{N}$ participants receive the latest global model $\theta_t$. Then, every participant $\it{k}$ trains the model by performing multiple iterations of stochastic gradient descent (SGD) on minibatches from its local dataset $D_k$. The local training results in a several weight-update vectors $\Delta \theta_{k}^{t+1}$, which are sent to the server. The last step is the model aggregation at the server, which is typically achieved using weighted aggregation \cite{r8} following Eq.\ref{eq:fedavg}. The process is then repeated until the model converges.
\begin{equation}
    \theta_{t+1} = \theta_{t} + \sum_{k=1}^{N} \frac{\left | D_k \right |}{\left | D \right |}\Delta \theta_{k}^{t+1}
\label{eq:fedavg}
\end{equation}

While this aggregation method takes into account the unbalanced aspect of datasets' size, it is not always suitable for non-i.i.d distributions. Furthermore, FL in wireless networks in general, and in vehicular networks in particular, is subject to the following challenges:\\
\textbf{Statistical heterogeneity:}
One of the underlying challenges for training a single joint model in FL settings is the presence of non-i.i.d data.  For instance, some nodes only have access to data from a subset of all possible labels for a given task, while other nodes may have access to different input features.
Furthermore, varying preferences for instance can lead to concept shift (i.e., nodes classify same features under different labels, or vice-versa). In practice, these non-i.i.d settings are highly likely to be present in a given massively distributed dataset. Thus, training models under these settings requires new sets of considerations.  

\textbf{Partial Participation:}
Given the scarcity of the communication resources, the number of participating nodes is limited. In fact, the generated traffic grows linearly with the number of participating nodes and the model size.
Moreover, the heterogeneity of the nodes in terms of computational capabilities and mobility (i.e., velocity and direction) introduces stringent constraints on the communication.
Hence, enabling FL on the road in a communication-efficient way is far from an easy task. 
\subsection{Related Work}
Several works consider FL as a key enabler for vehicular networks in general, and AD in particular \cite{yang_edge_2021}, such as secure data sharing \cite{lu_blockchain_2020}, Autonomous Controllers \cite{zeng_federated_2021}, caching \cite{yu_mobility-aware_2020}, and travel mode identification from non-i.i.d GPS trajectories \cite{gps_fl}. Nonetheless, deploying FL on the road remains a challenging task due to uncertainties related to mobility and communication overhead. To  overcome the communication bottleneck, works {\revision \cite{lim_federated_2020, aledhari_federated_2020,imteaj_survey_2021}} have proposed judicious node selection and resource allocation for efficient training. However, these schemes are specifically designed for the topology and dynamics of standard wireless/cellular networks with high node density but relatively low mobility. In contrast, vehicular networks have rather low node density and very high node mobility \cite{elbir_federated_2020}. As a result, new schemes are required for FL on the road. 
Meanwhile, V2V communication offers a new possibility for FL deployment that bypasses the bottleneck of communication with the MEC server\cite{meysam4,meysam5}. In vehicular networks, some vehicles serve as edge nodes to which neighboring nodes offload computation and data analysis tasks \cite{jabri_vehicular_2019}. Edge vehicles are also used to provide a gateway functionality by ensuring continuous availability of diversified services such as multimedia content sharing \cite{al_ridhawi_continuous_2018}. A common practice among such works is creating clusters of vehicles where the edge vehicle acts as a cluster head. The clusters are formed based on several metrics such as the distance between the vehicles, their velocity and direction. Yet, these clustering schemes cannot be directly exploited in the context of FL. {\revision Recent VANET clustering works principally design algorithms  based on their primary application \cite{tal_towards_2014,cooper_comparative_2017,meysam1,meysam2}. This is a logical approach since the design of a clustering algorithm highly influences the performance of the application for which it is used. 
A popular approach for cluster head selection widely used in the literature \cite{cooper_comparative_2017, singh_nwca_2016, daeinabi_vwca_2011} requires each vehicle to calculate an index quantifying its fitness to act as a cluster head for its neighbours. Vehicles wishing to affiliate with a cluster head rank all neighbours in their neighbour table and request association with the most highly-ranked candidate node.  The index is calculated as a weighted sum of several metrics, such as the degree of connectivity and link stability, with weights chosen depending on the importance of the considered metrics. However, due to the nature of FL applications, metrics related to learning/data should also be considered. 
}

Furthermore, clustering is already used in FL as a means to accelerate the training by grouping nodes with similar optimization goals, which train different versions of the model instead of one global model \cite{sattler_clustered_2020,kim_dynamic_2020,briggs_federated_2020,ghosh_efficient_2021}. In fact, one of the fundamental challenges in FL is the presence of non-i.i.d and unbalanced data distributions {\revision \cite{ kairouz_advances_2021,tak_federated_2021}}. These challenges go against the premise of FL which aims to train one global model. Such settings require new mechanisms to be put in place in order to ensure models' convergence. Clustered FL has attracted several research efforts, as it has generalization \cite{mansour_three_2020} and convergence \cite{ghosh_efficient_2021} guarantees under non-i.i.d settings. {\revision By creating different models to adapt to different end users' distributions, clustered FL allows better model performance in the case of concept-shift. Concept-shift \cite{briggs_federated_2020} occurs when different inputs do not have the same label across users as preferences vary. Moreover, in clustered FL, training becomes resilient to poisoning attacks \cite{chen_zero_2021} such as label flipping \cite{taik_data-quality_2021} (i.e., nodes misclassify some inputs under erroneous labels). }

For instance, authors in \cite{sattler_clustered_2020},  develop a clustered FL procedure. Their work allows to find an optimal bipartitioning of the users based on cosine similarity for the purpose of producing personalized models for each cluster. The bipartitioning is repeated whenever FL has converged to a stationary point. 
In \cite{briggs_federated_2020}, a single clustering step, in a predetermined communication round, is introduced. In this step, all the users are required to participate and the similarity of the updates is used to form clusters using hierarchical clustering. Nonetheless, the proposed approach requires knowing a distance threshold on the similarity values between the updates to form the clusters. Furthermore, cluster-based approaches assume that all the users participate, which is unfeasible under dynamic and uncertain vehicular networks. 

To the best of our knowledge, our work is the first to address the problem of clustered FL in hierarchical mobile architectures, while considering the users’ data distributions, wireless communication characteristics, and resource allocation constraints. Specifically, unlike other studies, we consider the learning aspect (i.e., nodes dataset characteristics and model dissimilarities),
in addition to communication constraints (i.e., wireless channel quality, mobility, and communication latency). Henceforth, we propose a practical way to deploy FL in vehicular environments.
\begin{figure*}[t]
	\centering
	\includegraphics[scale=0.45]{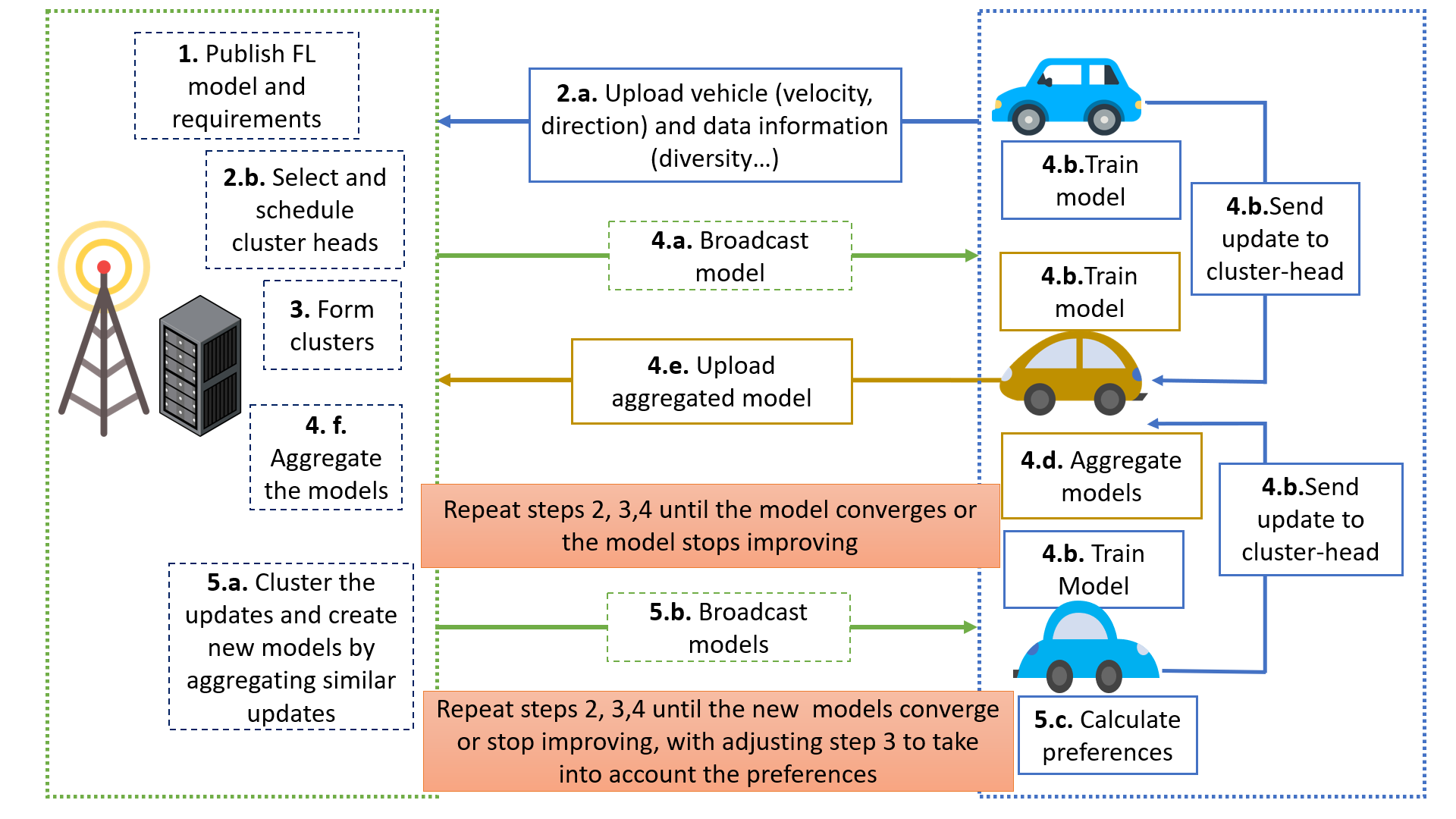}
	\caption{{\revision Illustration of the different steps in clustered vehicular federated learning}}
	\label{fig:archi}
\end{figure*}

%% file: 3_model.tex

 \section{System Model}
 
     We consider a vehicular network composed of a set $V$ of $K$ vehicles and a set $U$ of $N$ gNodeBs. Both the communication among vehicles and with the gNodeBs  are through wireless links. Additionally, gNodeBs are connected to the Internet via a reliable backhaul link. The vehicles have enough computing and storage resources for the training, and the gNodeBs are equipped with MEC servers. MEC servers are used to schedule the vehicles nearby, aggregate the updates and manage the clusters. In the following, we explain the proposed cluster-based training process and the different components of the considered system model (i.e., communication and computation) in a vehicular environment.

{\revision \subsection{Process Overview}
{\revision FL in vehicular networks is subject to several challenges related to data, mobility, and communication and computation resources. In this paper, we consider these aspects in the design and optimization of the FL process in vehicular networks. 

The first set of challenges are related to data, where the learning process should be adapted to take into account  data heterogeneity in order to accelerate the model convergence. 
Data generated across different applications in vehicular networks depend on the specific vehicle sensors and these sensors' data acquisition activities which often leads to heterogeneous data distributions among FL participants (i.e., different dataset sizes and different data distributions). Furthermore, the dependence on data acquisition activities from vehicles with similar sensing capabilities makes the collected data highly redundant. As a result, local datasets cannot be regarded the same in terms of information richness, as some datasets may have more diverse and larger datasets than other participants. Furthermore, communication resources in this context are limited. In fact, in addition to the bandwidth’s scarcity, the possible time for communication with the MEC server is limited by the time where a vehicle is in the area covered by the base station. For all these reasons, the participant selection and the bandwidth allocation mechanisms should be carefully designed for FL in vehicular networks. 
 Hence, in this article, we use the data properties to guide the participants' selection in the training and communication process.

Furthermore, the model convergence speed is highly dependent on the number of collected updates. Vehicle-to-vehicle (V2V) communication offers a great alternative to bypass the communication bottleneck in vehicular networks by allowing some select vehicles to act as mediators between other vehicles and the MEC server. We propose to use V2V in order to maximize the collected updates under the communication uncertainty. 

In these perspectives, we propose to prioritize the vehicles with the most informative datasets and use them as cluster heads, while the remainder of the vehicles are associated with them. 
In this setting, each cluster-head aggregates the models of the vehicles in its cluster and uploads the resulting model. In fact, instead of sending all the collected updates, the cluster-head will aggregate the updates and send one aggregated model which is more communication-efficient. In this case, hierarchical FL is used as a means to optimize the communication in vehicular networks, where the MEC server will do a second round of aggregation. 

Another aspect that needs to be considered is  mobility and how it affects the communication among vehicles and with the MEC server. In order for the cluster-heads to successfully upload their models to the MEC server, the upload should be completed before the vehicles leave the coverage area of the BS. Furthermore, for the vehicles to be able to send their models to the cluster-head, their link lifetime (LLT)  should be longer than the required time for training and uploading the models.

In order to adapt this approach to the case where multiple models need to be trained, other considerations need to be taken into account in this approach. In fact, in the case where data distributions are subject to concept-shift, a single model is not enough. Concept-shift is another kind of data heterogeneity that arises in cases where data is subjective and depends on the preferences of end users, or in the presence of adversaries. In classification problems for instance, concept shift is when similar inputs have different labels depending on the end user. In the case of vehicles, the latter could simply not share the same model if they are not from the same OEM.  
The presence of different perspectives from different vehicles makes one model hard to fit all.  In our paper, we use hierarchical clustering through evaluating the model updates and their cosine similarity. The clustering can be executed on a predetermined communication round or when the model’s convergence slows down. The newly created models will be used to associate each vehicle to the most adequate cluster-head. The same model can be trained among several clusters as such redundancy is worthwhile when it comes to system robustness in the case of user dropout, and it also helps the model's convergence through collecting more updates. 

All in all, to address the challenges linked to mobility and data heterogeneity, we design a mobility-aware scheme for clustered FL, that takes into account the data and model heterogeneity. The data heterogeneity is mainly considered in the selection of cluster-heads, while the model heterogeneity is used to create new models and in matching vehicles to cluster-heads. In the following subsections, we start with detailing the overall learning model, then we present the mathematical formulation of its different aspects.  We detail the steps of the clustered vehicular FL training procedure, then we give the formulations of the different metrics used in the procedure.

}

 }
\subsection{Learning Model}
 A summary of the process is given in Algorithm \ref{alg:procedure}, and more details of the scheme are given as follows:
\begin{itemize}
    \item \textbf{Step 1} (Publish FL model and requirements, and receive feedback ) : A global model is published by the MEC server, alongside its data and computation resource requirements (e.g., data types, data sizes, and CPU cycles).  Each vehicle $\it{k}$ satisfying the requirements sends positive feedback, in addition to other information such as its data diversity index $I_k$ {\revision (see Eq. \ref{eq:diversity})} and current velocity $v_k$.
    \item \textbf{Step 2} (Select and schedule cluster-heads $\it{H}$): The MEC server chooses the cluster-heads according to the received information. The selection is based on {\revision the dataset characteristics (i.e., quality of the dataset and the quantity of the samples), defined in subsection \ref{subsec:diversity},} in addition to the state of the wireless channel and the projected duration of the communication reflected by the rate of stay {\revision (See Eq. \ref{eq:stay})}. In fact, the quality of local dataset directly determines the quality and the importance of model updates, while the velocity and the state of the wireless channels determine whether the model update can be received during the communication round. {\revision The details about the data evaluation are given in subsection \ref{subsec:diversity} , and the algorithm (Algorithm  \ref{alg:schedule}) is explained in Section \ref{sec:algorithms}.}
    \item \textbf{Step 3} (Clusters formation): After cluster-head selection, the set of the remaining vehicles $\it{NH}$ are matched to cluster-heads (set $\it{H}$). The matching requires that the sum of training and upload time of vehicle $\it{k}$ is less than the Link Lifetime (LLT) {\revision (defined in Eq. \ref{eq:LLT})} between $\it{k}$ and $\it{h} \in \it{H}$ if they are to be matched.  Furthermore, the matching aims to maximize the weighted sum of $R_{k,h}$. $R_{k,h}$  symbolizes the relationship between $\it{k}$ and $\it{h}$, and its definition changes depending on whether there is only one global model or several versions {\revision (See Eq.\ref{eq:relationship})}. In the simple case of a single joint model, the clustering depends only on the mobility and accordingly for all the pairs $k \in NH , h \in H$ the value of $R_{k,h} = 1$. Otherwise, each vehicle should train its preferred model. The preference is defined as the accuracy of the model trained by $\it{h}$ on the local data of $\it{k}$. This definition is due to the fact that not all vehicles can participate in the updates clustering step (See Step 5). 
    \item \textbf{Step 4} (Model broadcast and training) :  The model is broadcasted to the participants, where each vehicle trains on its local data for $\epsilon$ local epochs, before sending the update to the corresponding cluster-head. Each cluster-head then aggregates the received models and sends the update to the MEC server, which in its turn aggregates the global updates of the clusters. Such hierarchical FL aggregation is widely adopted in the literature of FL \cite{liu_client-edge-cloud_2020,chai_hierarchical_2021} and allows for more participation. {\revision Aggregating the updates at the MEC server level is required because each model version can be trained within several clusters, resulting in several global models. Such redundancy is necessary in the case of vehicular networks, as it allows more robustness to client drop-out.}   
    \item \textbf{Step 5} (Updates Clustering and Preference Evaluation): If the global model does not converge after several communication rounds, or the goal accuracy is not attained, we perform a communication round (or several communication rounds) involving a large fraction of the vehicles on the global joint model. This step requires the collection of the updates at the MEC server without prior aggregation by cluster-heads {\revision as the aggregated models would mask the divergence of the different models}. The updates are used to judge the similarity {\revision (defined in Eq.\ref{eq:cosine})} between participants using the hierarchical clustering algorithm. It is employed to iteratively merge the most similar clusters of participants up to a maximum number of clusters defined by the OEM. Fixing the maximum number of clusters allows to create clusters without prior knowledge of the possible distances between updates, while controlling the number of models in circulation. Once the clusters are created, new models are generated through aggregation. The models are broadcasted to the available vehicles. Each vehicle evaluates the models on its local data and send them back to the MEC server. These values are later used to evaluate $R_{k,h}$ for each vehicle $\it{k}$. The resulting models are then trained independently but simultaneously using the same process. This preferences' evaluation makes the difference between our work and previous work in clustered FL, as these works necessitate the participation of all the nodes, while in our work we tolerate partial participation.
    
 \end{itemize}
\begin{algorithm}[t]
	\begin{algorithmic}[1]
	\For{$\it{i} \in [1 \dots i_{max}]$}
	    \If{$i=1$}\\
	    {\revision \textbf{Step 1: }}
	        \State initialize or download the newest model's parameters at the MEC server
	        \State initialize the number of models with 1
	        \State {\revision Publish model and training requirements}
	    \EndIf
	    \State {\revision \textbf{Step 2:}} Receive vehicles information (transmit power, available data size, dataset diversity, CSI, velocity, preferred model)
		\State Schedule cluster-heads $H$ using Algorithm 1 
		\State {\revision \textbf{Step 3:}} Assign the remainder of vehicles (i.e., $NH$) to clusters using Algorithm 2\\
		{\revision \textbf{Step 4:}}	
        \For {\text{ vehicle } $k \in NH$ }
			 \State $\it{k}$ receives model $\it{\theta_t}$
			 \State $\it{k}$ trains on local data $D_{k}$ for $\epsilon$ epochs
			\State $\it{k}$ sends updated model $\theta_k^{t+1}$ to MEC server
		\EndFor

		\For {\text{ cluster head } $h \in H$ }
			 \State $\it{h}$ trains on local data $D_{h}$ for $\epsilon$ epochs			 
			 \State $\it{h}$ receives model updates from vehicles in its cluster 

			\State $\it{h}$ aggregates the model and sends new global model to MEC server
		\EndFor\\
		{\revision \textbf{Step 5:}}
	    \If{$i=t_c$}
	    \State At step $i=t_c$ MEC server evaluates the similarities of the received models 
		\State MEC server creates clusters based on the similarities and computes new global models using weighted average
		\State nodes receive new global models and evaluate their preferences
	    \EndIf	
	    \State aggregate updates
        \State start next round $\it{i}\leftarrow \it{i}+1$		
	\EndFor
	\end{algorithmic}
	\caption{Clustered Vehicular Training procedure}
	\label{alg:procedure}
\end{algorithm}
\vspace{0.2cm}

{\revision The iterations and the steps' order are illustrated in Fig.\ref{fig:archi}.
Next, we present the formulations of the different elements in the system model, starting with the learning aspects (i.e., dataset charasteristics and models similarity), to the different mobility and communication aspects considered throughout the proposed approach.}
\subsubsection{Dataset characteristics}
\label{subsec:diversity}

Considering the fact that datasets are non-i.i.d and unbalanced, a judicious cluster-head selection {\revision (Step 2)} is necessary. 
In fact, each dataset can be characterized by how diverse its elements are, its size and how many times the model was trained on it (i.e., age of update). In this paper, we focus on the non-i.i.d and unbalanced aspect, however, other metrics can be considered depending on the learned task, including the quality of the datasets and their reliability.
We set the value of each metric as \cite{taik_data-aware_2021}: $ \varphi_j \gamma_j$,  where  $\gamma_j$ is the adjustable weight for each metric assigned by the server and $\varphi_j$ is the normalized value of the metric $\it{j}$. Using the aforementioned characteristics, the diversity index of dataset at node $k$ can be defined as: 
\begin{equation}
I_k = \sum_{j} \varphi_{j,k}\gamma_{j},
\label{eq:diversity}
\end{equation}
with $j \in\{\text{elements diversity}, \text{dataset size}, \text{age}\}$. The metric can be easily adjusted to include other task-specific considerations.

\subsubsection{Updates similarity}
In order to handle the non-i.i.d aspect, the updates' similarity is evaluated using cosine similarity \cite{sattler_clustered_2020,briggs_federated_2020} {\revision in Step 5 of the algorithm}, and new models are created by aggregating the most similar models. Given two model updates $\Delta \theta_k$ and $\Delta \theta_l$, the similarity is calculated according to: 
\begin{equation}
    sim(k,l) = \frac{ \left \langle \Delta \theta_k,\Delta \theta_l \right \rangle}{\left \| \theta_k \right \|\left \| \theta_l \right \|} 
    \label{eq:cosine}
\end{equation}
{\revision where $\left \langle . , .  \right \rangle$ is the dot product of two vectors. The dot product is divided by the product of the two vectors' lengths (or magnitudes).}
The values of $sim(.)$ are between 0 and 1, and the dissimilarity (i.e., cosine distance metric) $1-sim(.)$ is used to cluster the updates. The cosine distance metric is invariant to scaling effects and therefore indicates how closely two vectors (and in our case updates) point in the same direction. 
{\revision The models' similarity is then used to created clusters using the hierarchical clustering algorithm \cite{briggs_federated_2020}, and the most similar models are aggregated to create new models.}

\subsubsection{Vehicles Relationships}
{\revision During the cluster formation in Step 3, each cluster is created based on the relationship between the vehicles. The definition of this relationship depends on whether only one global model is trained, or there are several versions of the model that are created. In the case of multiple models, we define the preference of a model through its accuracy on the $k$th vehicle's dataset.}
We define the relationship between two vehicles $R_{k,h}$ as follows:
\begin{equation}
 R_{k,h}   =
\left\{\begin{matrix}
 \text{accuracy of $h$} & \text{if more than 1 model}\\ 
 1 & \text{otherwise} 
\end{matrix}\right.
\label{eq:relationship}
\end{equation}

\subsection{Communication Model}
In Step 2, due to mobility and communication constraints, the RB allocation is jointly executed with the cluster-head selection. In fact, the mobility imposes a deadline for the upload based on the standing time of the vehicle. Additionally, in Step 3, the cluster formation must also consider the relationship between the vehicles in terms of mobility, which is modelled through the link lifetime (LLT). The different aspects of the communication model are formulated as follows:  
\subsubsection{Standing time}
While typically in FL, the duration of a communication round is fixed by the centralized entity (e.g., MEC server), the latency in FL in vehicular networks is dictated by the standing time of participating nodes. 
Let the diameter of coverage area of a gNodeB be denoted as $\it{D}$. For
each vehicle $\it{k}$, the standing time in the coverage area of current
gNodeB is defined by Eq. \ref{eq:stay} \cite{yu_mobility-aware_2020}:
 
\begin{equation}
  T_k = \frac{D - x_k}{v_k} 
  \label{eq:stay}
\end{equation}
To ensure the communication with the gNodeB, the rate of standing time of a vehicle $\it{k}$ selected as cluster-head should respect $(t_{k}^{train}+t_{k}^{up}+T_{agg}+\delta)\leq T_k$. Where $t_{k}^{train}$ and $t_{k}^{up}$ are the estimated training time and upload time of vehicled $k$ respectively, $T_{agg}$ is the time required for aggregation and $\delta$ is a waiting time for the updates' collection. We can notice that what varies the most among the vehicles are $t_{k}^{train}$ and  $t_{k}^{up}$, as $t_{k}^{train}$ depends on the size of the dataset, and $t_{k}^{up}$ depends on the channel gain and the resource block allocation.

\subsubsection{Resource Blocks}
For each vehicle $\it{k}$, we can infer the maximum $t_{k}^{up}$ by setting $(t_{k}^{train}+t_{k}^{up}+T_{agg}+\delta) = T_k$. As a result, we can determine the minimum required data rate $r_{k,min}$ to send an update of size $\it{s}$ within a transmission time of $t_{k}^{up}$ as follows:
\begin{equation}
t_{k}^{up} = \frac{s}{r_{k,min}}.    
\end{equation}
The achievable data rate of a node $\it{k}$ over the RB $\it{q}$ is defined as follows:
\begin{equation}
  r_k^q = B \log_2(1 + \frac{P_k G_{k,q}}{N_0})
  \label{eq:datarate}
\end{equation}
where $\it{B}$ is the bandwidth of a RB, $P_{k}$ is the transmit power of node $\it{k}$, and $N_{0}$ is the power spectral density of the Gaussian noise. The data rate of a vehicle is the sum of the datarates on all the RBs assigned to it. 
\subsubsection{Link Lifetime}
{\revision In Step 3,} in order to associate a vehicle $k \in NH$ to a cluster-head $h \in H$, it is necessary to evaluate the sustainability of the communication link, so as to ensure that the update of the node $\it{k}$ will be successfully sent to $\it{h}$. 
Link Lifetime (LLT) \cite{ren_unified_2018} defines the link sustainability as the duration of time where two vehicles remain connected. 
{\revision LLT is defined in \cite{ren_unified_2018,li_robust_2012,ren_mobility-based_2017}} by Eq. \ref{eq:LLT}, for two vehicles $\it{k}$ and $\it{h}$ moving in the same or opposite directions. Assuming that the trajectory of all vehicular nodes to be a straight line, as the lane width is small, the y-coordinate can be ignored. We denote the positions of $\it{k}$ and $\it{h}$ by $x_k$ and $x_h$ , respectively.
\begin{equation}
  LLT_{k,h} =  \frac{-\Delta v_{kh} \times D_{kh} + \left | \Delta v_{kh}  \right | \times TR}{(\Delta v_{kh})^{2}} 
  \label{eq:LLT}
\end{equation}
with $\Delta v_{kh} = v_k - v_h$ and $D_{kh}= x_k - x_h$ and $\it{TR}$ denotes the transmission range.
Accordingly, the training time of $\it{k}$ and upload time from $\it{k}$ to $\it{h}$ must be less or equal to $LLT_{kh}$ (i.e., $(t_{k}^{train}+t_{k}^{up, h}) \leq LLT_{kh}$. 

%% file: 4_design.tex
\section{Problem formulation \& Proposed Solution}

\subsection{Problem Formulation}

Considering the collaborative aspect of FL and the communication bottleneck, we define the following goals for the cluster-head selection and cluster association: 
\begin{itemize}
    \item From the perspective of accelerating learning and maximizing the representation, the scheduled cluster-heads must have diverse and large datasets, as a result the goal of cluster-head selection is: 
    \begin{equation}
        \max_{h,\alpha} \sum_{k=1}^{K}{h_{k}I_{k}}.
    \end{equation}
    \item In order to guarantee that each vehicle trains its preferred model, the cluster assignment can be defined as a matching problem where we aim to maximize the relationship $R_{k,h}$. 
    \begin{equation}
        \max_{m} \sum_{h\in H}\sum_{v\in NH} R_{v,h} m_{v,h} 
    \end{equation}
\end{itemize}

Several constraints related to communication are imposed by the vehicular environment. Consequently, the first problem considered is a joint cluster-head selection and RB allocation. For each vehicle $\it{k}$ and RB $\it{q}$ we define $\alpha_{k,q}$ as: 
\begin{equation}
 \alpha_{k,q}   =
\left\{\begin{matrix}
 1& \text{if $q$ is assigned to k}\\ 
 0& \text{otherwise} 
\end{matrix}\right.
\end{equation}

The cluster-head selection and RB allocation problem is formulated as follows: 
\begin{maxi!}|c|[3]
 {h,\alpha}{{\sum_{k=1}^{K}{h_{k}I_{k}}}}
 {}{} \label{pbmoo}
\addConstraint{\qquad (t_{k}^{train}+\delta+ t_{k}^{up}+T_{agg}) h_k \leq T_k, \qquad \forall k \in [1,K]}\label{eq:TrainUploadT}
\addConstraint{\qquad \sum_{k=1}^{K}\alpha_k  \leq Total_{RB},\qquad  \forall k \in [1,K]}\label{eq:Bandwidth}
\addConstraint{\qquad h_k \in \{0,1\}, \qquad \forall k \in [1,K].}\label{eq:x_bounds}
\end{maxi!}


Taking into account the results from the previous problem, we define $H= \{ k, h_k = 1 \}$ (i.e., the cluster-heads) and $NH = \{k, h_k = 0 \}$ (i.e., the remainder of the vehicles). 
The next step is matching the set of vehicles $NH$ to selected cluster-heads $H$. We consider that a maximum capacity $N_{max}$ is fixed for each cluster in order to reasonably allocate the V2V communication resources. Additionally, if a vehicle $\it{v}$ is to be matched with a cluster-head, it needs to respect the time constraints, where it should be able to finish training and uploading before a deadline $T_{h} = t_{h}^{train}+\delta$, and the $LLT{v,h}$ should at least outlast the training and upload. We define $m_{v,h}$ as a binary variable equal to 1 if $v$ is matched with $h$ and 0 otherwise. Accordingly, we define the second problem as follows:
 
\begin{maxi!}|c|[3]
 {m}{{\sum_{h\in H}\sum_{v\in NH} R_{v,h} m_{v,h} }}
 {}{} \label{pbmoo2}
\addConstraint{\qquad \sum_{h \in H} m_{v,h}  \leq 1,\qquad  \forall v \in NS}\label{eq:match1Cluster}
\addConstraint{\qquad \sum_{v \in NH} m_{v,h}  \leq N_{max},\qquad  \forall v \in NS}\label{eq:matchNCluster}
\addConstraint{\qquad (t_{v}^{train}+t_{v}^{up}) m_v,h \leq LLT_{v,h}, \qquad \forall v \in NH}\label{eq:TrainUpCluster}
\addConstraint{\qquad (t_{v}^{train}+t_{v}^{up}) m_v,h \leq T_{h}, \qquad \forall v \in NH}\label{eq:TrainUpCluster2}
\addConstraint{\qquad m_{v,h} \in \{0,1\}, \qquad \forall v \in NH.}\label{eq:x_bounds2}
\end{maxi!}

\label{sec:algorithms}
\subsection{Proposed Algorithm}
In this section, we present our proposed solution for cluster-head selection and RB allocation alongside the matching algorithm to solve (12) and (13).
{\revision The challenging aspect of the problem (12) is that it requires maximizing the weighted sum of the selected vehicles and jointly allocating the bandwidth. A restricted version of problem (12) can be shown to be equivalent to a knapsack problem and thus it is NP-hard \cite{pisinger_where_2005}. In fact, the problem aims to select vehicles that maximize the weighted sum $\sum_{k} I_k h_k$ subject to a knapsack capacity given by $\sum_{k} \alpha_k \leq Total_{RB}$ in constraint (12c), which can be transformed to $\sum_{k} \alpha_kh_k \leq Total_{RB}$ where $\alpha_k$ represent the weight of item $k$ (fixed for this restricted version) and $Total_{RB}$ represents the knapsack capacity. Thus, the problem is equivalent to a knapsack problem and since the latter is NP-hard, so is problem (12). Constraint (12b) can be verified for each vehicle to filter out the ones that cannot upload the updates in time.}

We chose to follow a greedy knapsack algorithm to solve the problem. 
In fact, we chose the greedy approach because it will allow us to select the best candidates with an optimal RB cost, unlike the ranked list solution,
which would have optimized the sum of $I_k$ only \cite{cormen_introduction_2009}. Furthermore, the greedy knapsack algorithm has low complexity and will allow fast and efficient scheduling under the rapidly changing vehicular environment.
We calculate the minimum required RBs for each vehicle $\it{k}$ to be able to send the update by the deadline $T_k$, which we consider the cost of the scheduling $c_k = \sum_{q \in RBs} \alpha_{k,q}$. {\revision The main time consuming step is the sorting of all vehicles in a decreasing order based on their diversity value / cost in RBs ratio. After the vehicles are arranged as an ordered list, the following loop takes $O(n)$ time. Taking into account that the worst-case time complexity of sorting can is $O(n\log n)$, the total time complexity of the proposed greedy algorithm is $O(n\log n)$.}

\begin{figure}[t]
	\centering
	\includegraphics[width=8.5cm, height=6.2cm]{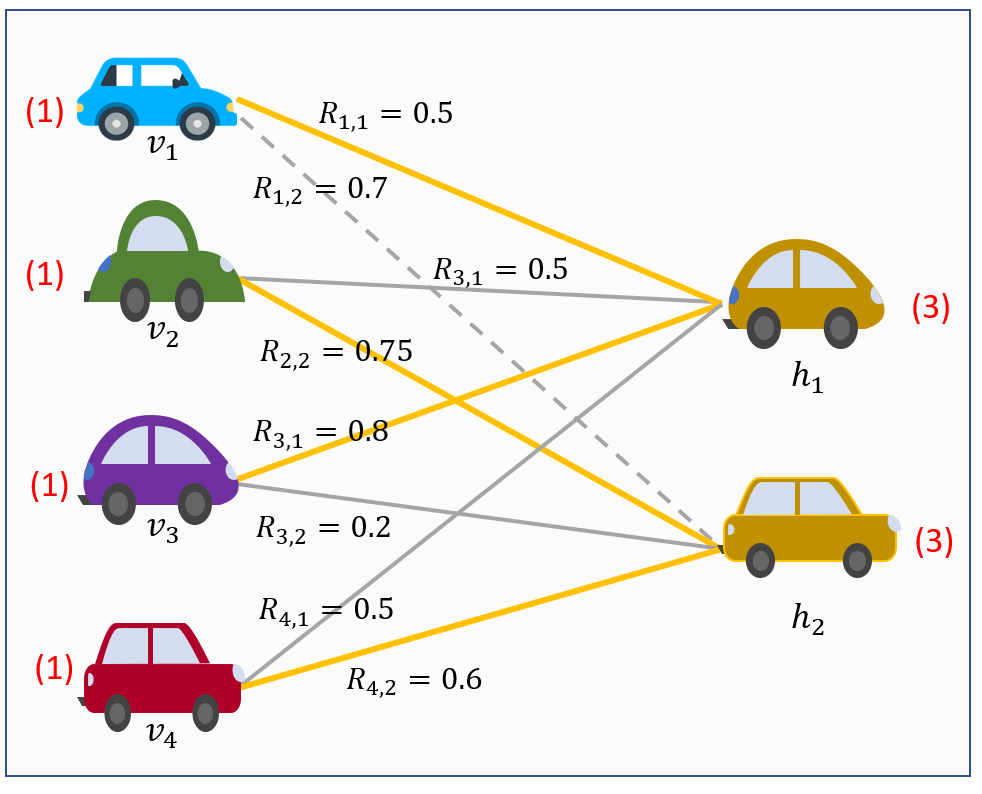}
	\caption{{\revision Illustration of the Matching problem} }
	\label{fig:matchi}
\end{figure}

{\revision The second formulated problem (13) is a maximum weighted bipartite matching problem \cite{chen_group-aware_2016,matching2017}, where each $h \in H$ has a maximum capacity $N_{max}$ and each $v \in NH$ has a capacity of 1. 


In order to include the remainder of the constraints, we define $\zeta_{v,h}$ as a binary value, where $\zeta_{v,h} = 0$ if constraint (13d) cannot be satisfied if $m_{v,h} = 1$, and  $\zeta_{v,h} = 1$ otherwise. The goal is redefined so as to  maximize a weighted sum of $R_{v,h} \times \zeta_{v,h}$. The problem becomes an integer linear program (ILP) and solved using an off-the-shelf ILP solver (e.g., Python's PulP \cite{noauthor_optimization_nodate}). 

To illustrate the problem, we consider the example in Fig.\ref{fig:matchi}. The vehicles and their relationships can be considered as a graph, where the vehicles represent the edges and their relationship is represented through the vertices, which are weighted with $R_{v,h} \times \zeta_{v,h}$. The goal is to find a subgraph where the selected vertices have an optimal (in our case maximum) sum. The remaining constraints are the maximum capacities of the vehicles (in red). The cluster-heads (in yellow, on the right) have a maximum capacity $N_{max} = 3$ each (Constraint \ref{eq:matchNCluster}), and the other vehicles have capacity of $1$ (Constraint \ref{eq:match1Cluster}). 
In the illustrated problem, the pairs $v_2, h_2$ and $v_3 , h_1$ cannot be matched since the edges ( in dashes lines ) have null values, which can be either due to possible disconnection or of poor model performance. The choice of the optimal matching is then left among the remaining pairs. The optimal solution for the illustrated problem in yellow lines has a sum of $3.0$. 
}
 
We define our Algorithm \ref{alg:schedule}, Clustered Vehicular FL (CVFL) that iteratively selects nodes with best ratio $\frac{I_k}{c_k}$ to be cluster heads, and then matches the rest of the vehicles to them after verifying the time constraints by creating clusters that maximize $\sum_{h \in H}\sum_{v\in NH} R_{v,h} \times \zeta_{v,h}$. 

\begin{algorithm}[htb]
\hspace*{\algorithmicindent} \textbf{Input}  { A queue of $\it{K}$ vehicles
total available resource blocks $Total_{RB}$;}\\
	\hspace*{\algorithmicindent} \textbf{Output} { $\alpha$, $h = [h_1,\dots,h_K]$};
	\begin{algorithmic}[1]
	 \State  \textit{// Cost Evaluation}
	\For{$ k = 1 ,\dots, K$}
      \State  $r_k = 0, c = 1;$
      \State order the RBs using $r_k,q$;
      \While {$r_k\leq r_{k,min}$ and $c\leq Total_{RB} $}
        \State $q^* \leftarrow \arg \max_{q\in Z} G_{k,q}$;
        \State $r_k \leftarrow r_k + r_{k,q}$;
        \State $c \leftarrow c+1$;
    \State $c_k \leftarrow c$;
        \EndWhile 
    \EndFor\\
      \Return $C=[c_1 ,\dots, c_K]$  
    \State \textit{// RB Allocation}  
    \State{order vehicles according to their ratio ($L = [ \frac{I_k}{c_k} \forall k ]$) decreasingly;} 
    \For{$ k = 1 \dots K$}
         $h_k  \leftarrow 0 $; 
	\EndFor 
 	\State{$A \leftarrow Z$;}
 	\State{$k \leftarrow \arg\max(L)$;}
 	\While {$A \neq \o $}
     \State order the RBs using $r_k,q$;
      \While {$r_k\leq r_{k,min}$ and $c\leq Total_{RB} $}
        \State $q^* \leftarrow \arg \max_{q\in A} G_{k,q}$;
        \State $r_k \leftarrow r_k + r_{k,q}$;
        \State $ \alpha_{k,q}\leftarrow 1 $;
        \State $A \leftarrow A \setminus \{q\}$;        

        \EndWhile
    \State $h_k \leftarrow 1$;
    \EndWhile \\ 	    

    \Return $h$ and $\alpha$ 
    
    \State \textit{//  Matching}
    \State Use $h$ to form $H$ and $NH$ sets;
    \State Infer values of $R_{k,h} \forall k \in NH, h \in H$;
    \State Estimate $LLT_{k,h} \forall k \in NH, h \in H$;
    \State verify time constraints and calculate $\zeta$;
    \State Solve matching problem using Maximum weight bi-partite matching algorithm \cite{chen_group-aware_2016} using off the shelf solver such as Python's PulP \cite{noauthor_optimization_nodate}. 
    \State Uniformly allocate the RBs of V2V links to the associated vehicles.
	\end{algorithmic}
	\caption{Clustered Vehicular Federated Learning (CVFL)}
	\label{alg:schedule}
\end{algorithm}

%% file: 5_evaluation.tex
\section{Performance Evaluation}

\subsection{ Simulation Environment and Parameters}\label{subsec:param1}
The simulations were conducted on a desktop computer with a 2,6 GHz Intel i7 processor and 16 GB of memory and NVIDIA GeForce RTX 2070 Super graphic card. We used Pytorch \cite{noauthor_pytorch_nodate} for the machine learning library. In the following numerical results, each presented value is the average of multiple independent runs.

\vspace{0.2cm}

\vspace{0.2cm}
\textbf{Datasets:} We used benchmark image classification datasets MNIST \cite{mnist},a handwritten digit images, {\revision and Fashion-MNIST \cite{xiao_fashion-mnist_2017}, grayscale fashion products dataset}, which we distribute randomly among the simulated devices. 
MNIST and FashionMNIST constitute simple yet flexible tasks to test various clustered settings and data partitions. Each dataset contains 60,000 training examples and 10,000 test examples.
The data partition is designed specifically to illustrate various ways in which data distributions might differ between vehicles. The data partition we adopted is as follows: 
We first sort the data by digit label, then we form 1200 shards composed of 50 images each. Each shard is composed of images from one class, i.e. images of the same digit.  In the beginning of every simulation run, we randomly allocate a minimum of 1 shard and a maximum of 30 shards to each of the $\it{K}$ vehicles. This method of allocation allows us to create an unbalanced and non-i.i.d distribution of the dataset, which is varied in each independent run.\\
Furthermore, in order to evaluate the updates' clustering and how adequate is the preferences' evaluation,  we partition the vehicles' indexes into $N_{shifts}$ groups. For each group two digit labels are swapped. For instance, one group might swap all digits labelled as 1 to 7 and vice versa. The swapped tuples are: $\{(1,7),(3,5)\}$ for MNIST and $\{(1,3), (6,0)\}$ for FashionMNIST  \cite{tolpegin_data_2020}. Each group is then evenly distributed to $\frac{K}{N_{shifts}}$. This partition allows us to test the proposed algorithm’s ability to train models in the presence of concept shift and unbalanced data. The test set is divided into $N_{shifts}$ datasets and the average accuracy is then reported.\\

\textbf{FL Parameters:}\\
We consider $K = 30$ vehicles  collaboratively training 
{\revision multi-layer perceptron (MLP) model with two hidden layers (64 neurons in each), and a convolutional neural network (CNN) model with two 5x5 convolution layers (the first with 10 channels, the second with 20, each followed with 2x2 max pooling), two fully connected layers with 50 units and ReLu activation, and a final softmax output layer. We use lightweight models as they can be realistically trained on end-devices in rapidly changing environments.} For each participant, {\revision  due to the mobility of the vehicles and in order to collect a maximum number of updates, it is more practical to choose a small number of local epochs, as a result, in the following simulations, the number of local epochs is set to $\epsilon = 1 $.}  In the preliminary evaluations, the maximum number of communication rounds is $i_{max} = 30$. The clustering is set in round 25. $t_{k,train}$ for each vehicle is calculated locally using our configuration. 

\subsection{Preliminary evaluations: Parking Lot Scenario}
In this part of the evaluations, we focus on the learning aspect by studying the proposed algorithm in less constrained environment. 

\subsubsection{Simple unbalanced and non-i.i.d distribution}
\begin{figure}[htb]
	\centering
\includegraphics[width=8.7cm, height=6.3cm]{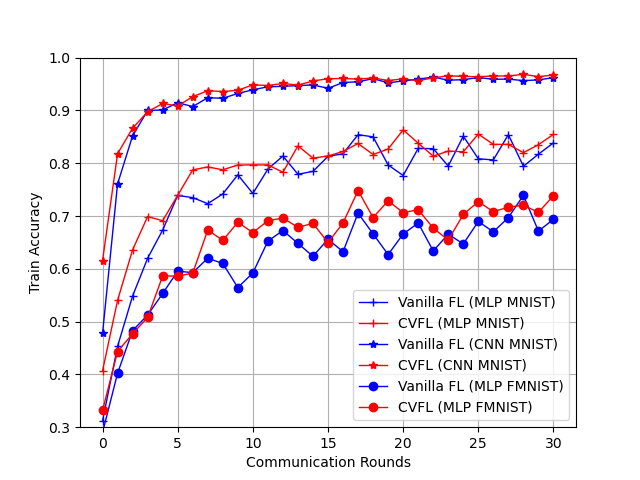}
	\caption{{\revision Preliminary results on non-i.i.d and unbalanced data without concept-shift}}
	\label{fig:prelim}
\end{figure}

In this part of the simulation, we ignore the constraint of LLT in problem (13) as the velocities are set to 0. The results in Fig.\ref{fig:prelim} show that a significant improvement is reached through the use of V2V communication. With more participation, we also noticed that the training tends to be more stable with the loss function steadily declining in comparison to standard FL. Furthermore, higher accuracy scores are achieved by our proposed method.{\revision While the average local accuracy after the end of the training the MLP on MNIST is $80\% \pm 10\%$for vanilla FL, it reaches and average of $82\% \pm 9\%$ for our proposed approach. Similarly, on FashionMNIST the results $66.79\% \pm 10\%$ with vanilla FL and $68.74\% \pm 9\%$. 
Owing to its high suitability for image processing tasks, the CNN model yielded higher results as the vanilla FL reached $94.58\% \pm 7\%$
and our proposed method achieved $95.5 \% \pm 5\%$.} Such results can be considered as a baseline values in perfect conditions for the subsequent experiments as we can reflect on the robustness of CVFL under mobility and concept-shift.
Based on these preliminary results, we expect to see more differences and variance in the results for the MLP model compared to the CNN model. We also can expect a better performance for the MLP model on the MNIST dataset compared to FashionMNIST. 
 
\vspace{0.3cm}
\subsubsection{Unbalanced and non-i.i.d distribution with concept shift}
The presence of concept-shift requires the clustering phase in order to improve the final results. In these simulations, we fixed the number of maximum clusters to 2, and studied the effect of partial participation on the clustering. Given the presence of concept shift for 4 out of 10 digits, we expect the accuracy to be around 60\%. 

\begin{figure*}[htb]
	\centering
\includegraphics[scale = 0.245]{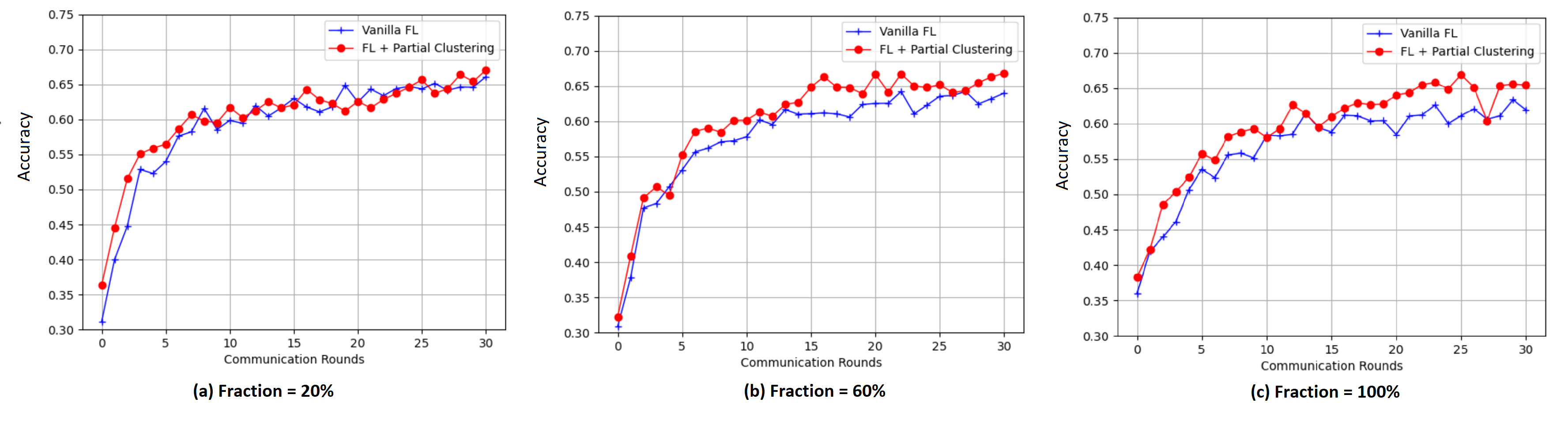}
	\caption{The importance of the fraction of the participants in the clustering step under concept-shift }
	\label{fig:fractions}
\end{figure*}

To study the effect of the fraction of participants in the partial clustering phase , we run multiple independent runs for each fraction in $\{20\%, 60\% , 100\%\}$. The results are shown in Fig.\ref{fig:fractions}. For both vanilla FL and the proposed partial clustering approach, the number of participants in each round is 6. For the standard FL, the average accuracy is 65\%, while 
For 20\% the average 68\% (+3\%) and for 60\% the average is 69\% (+5\%). It should be noted that the dissimilarity of the updates is harder to detect as only 2 out of 10 digits are swapped for each group. 

\begin{figure*}[htb]
	\centering
    \includegraphics[scale=0.25]{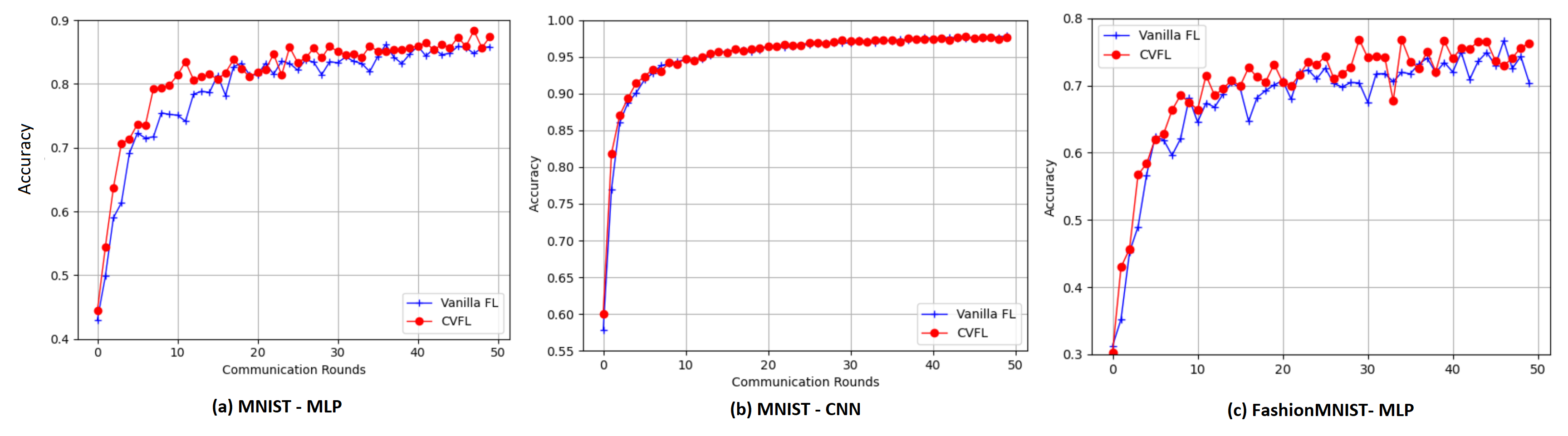}
	\caption{{\revision Evaluation of CVFL when the relationship is defined through mobility only}}
	\label{fig:freeway_mobonly}
\end{figure*}

\begin{figure*}[htb]
	\centering
    \includegraphics[scale=0.25]{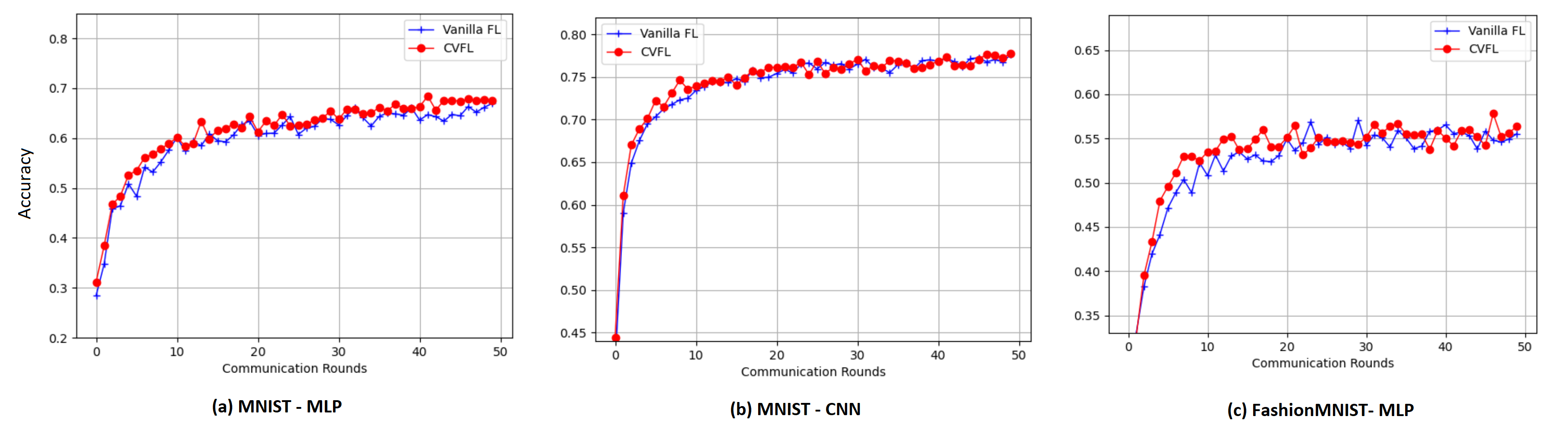}
	\caption{{\revision Evaluation of CVFL under concept shift}}
	\label{fig:freeway_mob_shift}
\end{figure*}

\begin{table}[h]
         	\centering
     		\caption{Generated Values}
     		\begin{tabular}[h]{|c|c|}
     		
     			\hline
     			Vehicle Antenna height & 1.5m \\ \hline 
                Vehicle antenna again & 3dBi \\ \hline
                Shadowing distribution & Log-normal \\ \hline
                Shadowing standard deviation & 3 db \\ \hline
                Noise power $N_0$ & -114 dBm \\ \hline
                Fast fading & Rayleigh fading \\ \hline
     			Transmit Power & 0.1 Watt     \\\hline
                Vehicles generation model & Spatial Poisson Process \\\hline
                Velocities generation model & Truncated Gaussian \\ \hline
     			Model Size & 160 kbits     \\ \hline
     			Bandwidth/ RB & 180 Khz    \\ \hline
     			$N_{max}$ & 2 \\ \hline
     			Total RBs & 4\\ \hline 
     			$\delta$ & 2s\\
     			\hline

     		\end{tabular}
     		\label{table:table2}

\end{table}

\subsection{Freeway scenario}
We consider that the $K = 30$ vehicles are randomly distributed on 6 lanes on a radius $D=2km$.
The vehicular communication model parameters and mobility are based on parameters in \cite{liang_resource_2017} and are summarized in Table \ref{table:table2}.
The velocities of vehicles are assumed to be i.i.d, and they are generated by a truncated Gaussian distribution. In contrast to the normal Gaussian distribution or constant values, the truncated Gaussian distribution is more realistic for modelling vehicles’ speed as it can generate different values in a certain limited range. This assumption is widely adopted in many state-of-the-art works of vehicular networks \cite{yu_mobility-aware_2020}. 
The lower and upper bounds for the velocity values on the 3 lanes going in the same direction are $(60,80),(80,100),(100,120) km/h$.
{\revision
\subsubsection{Key Performance results}
In this part of the evaluation, we vary the model size and the number of RBs in order to evaluate how the CVFL algorithms adapts to different training and upload requirements. We evaluated how the number of selected cluster-heads and how the total number of participants change in each scenario. We also evaluated how the average running time of the matching algorithm when the number of participants varies.

Table-\ref{table:nbrch} shows the average number of cluster-heads selected in each communication round and Table-\ref{table:nbrpart} shows the average number of participants in each communication round. It is clear from the results that the number of RBs is the defining factor of the number of cluster-heads and consequently the number of participants. The results also show that the proposed algorithm can safely scale up to handle large models or more local epochs in the case of small models. 

\begin{table}[h]
         	\centering
     		\caption{ {\revision Average Number of cluster heads in each communication round} }
     		\begin{tabular}[h]{|c|c|c|c|}
     			\hline
     			 &\multicolumn{3}{c|}{Model Size in Kbits}\\ \hline
     			 Number of RBs & 160 & 320 &640 \\ \hline
     			 2 & $2.43 \pm 0.26$ & $2.39 \pm 0.30$ & $2.39 \pm 0.24$ \\ \hline
     			 3 & $4.13\pm0.57$ & $4.056\pm0.51$ & $4.216\pm 0.48$ \\ \hline
     			 4 & $5.565 \pm 0.69$ & $5.504 \pm 0.71$ & $5.568 \pm0.54$ \\
     			\hline

     		\end{tabular}
     		\label{table:nbrch}
     
\end{table}     	

\begin{table}[h]
         	\centering
     		\caption{{\revision Average Number of participants in each run} }
     		\begin{tabular}[h]{|c|c|c|c|}
     			\hline
     			 &\multicolumn{3}{c|}{Model Size in Kbits}\\
     			 \hline
     			 Number of RBs & 160 & 320 &640 \\
     			 \hline
     			 2 &  $7.28 \pm 0.80$ &  $7.16 \pm 0.91$ & $7.168 \pm 0.74$ \\ \hline
     			 3 & $12.26\pm 12.05$ & $1.36\pm 0.51$ & $12.44\pm 1.39$ \\ 
     			 \hline
     			 4 & $16.00 \pm1.46$ &  $15.81 \pm1.47$&   $16.04 \pm 1.24$ \\
     			\hline

     		\end{tabular}
     		\label{table:nbrpart}
     
\end{table}  
\begin{table}[h]
     	\centering{
     		\caption{{\revision Average Running time of the matching algorithm}}
     		\begin{tabular}{|c|c|c|c|c|r|r|r|r|r|}
     			
     			\hline
     			Number of vehicles &  Average CPU time (s)    \\
     			\hline
     			 25   & 0.02     \\ \hline
     			 50   & 0.03   \\ \hline
     			 75  & 0.03     \\ \hline
     			 100   & 0.03   \\ \hline
     			 125   & 0.04  \\ \hline
     		\end{tabular}
     	
     	\label{tab:time}}
     \end{table}
}
{\revision Calculating the analytical expression of time complexity of the ILP-based algorithm used for the matching is not obvious since the low-level implementation details of the solver are not available to us. However, we evaluated the running  time in different settings with varying the number of nodes to see how it scales with large number of participants. The average running time values in seconds on our machine are summarized in Table-\ref{tab:time}. In general, the matching algorithm can easily handle large pools of participants without high impact on the execution time.}

\subsubsection{Effect on the accuracy}
To study the proposed approach in a mobility scenario, we first studied a simple case of unbalanced and non-i.i.d distribution, then we stress tested CVFL under concept-shift. The number of available RBs in each communication round is limited to 4, and the simulations were conducted for $i_{max} = 50$ communication rounds. 

Fig. \ref{fig:freeway_mobonly} shows the results for unbalanced and non-i.i.d distribution in the mobility scenario.

Owing to larger numbers of participants (see Table \ref{table:nbrpart}), higher accuracy values are obtained across the experimnents. CVFL achieves accuracy of $87\% \pm 4\%$ in contrast to $85\% \pm 5\%$ for the standard FL under the same settings training MLP model on MNIST, and the CNN model achieves similar results for both CVFL ($95\% \pm 5\%$) and vanilla FL ($94\% \pm 7.5\%$). The average accuracy values on FashionMNIST is $69.66\% \pm 9\%$ for CVFL and $66.46\% \pm 10\%$ for vanilla FL. 
The larger values of the standard deviation of the results in vanilla FL across the experiments in this case is possibly due to the smaller number of participants in each round compared to CVFL where almost half of the vehicles train their models which provides more consistency throughout the experiments.

\vspace{0.3cm}

The second set of simulation runs are on unbalanced and non-i.i.d distribution with concept shift. Fig.\ref{fig:freeway_mob_shift} shows how the models performed under these conditions in a freeway setting. Overall, accuracy values are significantly less than the obtained values in datasets where there is not concept shift. More specifically, {\revision the average accuracy of the MLP model achieved in the 50th round on MNIST dataset is  $68\%\pm 9 \%$} in contrast to {\revision $65\% \pm 7\%$ } for vanilla FL. The CNN model yielded identical results for CVFL ($80\% \pm 9\%$) and vanilla FL ($80\% \pm 5\%$). The larger values of the standard deviations in CVFL are due to the fact that resulting models after clustering often perform differently on the test sets.  
The concept-shift appears to affect the accuracy on FashionMNIST in a higher level, as the accuracy drops to around $55\%$ for both CVFL and vanilla FL. In contrast to the previous experiments, the difference is low in later rounds because only a small fraction of users participate in the clustering step. This can be overcome though the introduction of more communication rounds on the same version of the model in order to collect more updates. Additionally, the gaps in accuracy values are high in the earlier rounds of communication before the clustering round. The reason for the gap's narrowing is that new models are created and a smaller number of clients train the same model. As the number of clients training the models in each round constitutes a key factor for the convergence speed, we suspect that it might be the reason.




%

%% file: 6_conclusion.tex
{\revision
\section{Limitations and future work}
Through this work, we have identified several potential future research directions and open issues that are worthwhile being explored.
\begin{itemize}
    \item \textbf{Large-scale collaboration:} Extending the proposed model to take into account handover between base stations etc in order to enable continuous training throughout vehicles' trips and reduce lost updates. 
    Furthermore, fully decentralized training can be implemented for areas with low coverage, while also taking into consideration model convergence. 
    \item \textbf{Adversarial attacks and outliers:} The updates' clustering is useful to detect local models that diverge from the majority of the received updates. This step can be furthered exploited to eliminate outliers and adversaries. Additionally, due to the collaborative and hierarchical nature of the proposed approach, trust among vehicles and reliability of their models can be further enhanced through traceability and incentive/punishment mechanisms \cite{hajar_blockchain1}. 
    \item \textbf{Experimental values:}  Set thresholds concerning LLT and rate of stay through experimental/ real data traces. Other values related to training can also be adjusted dynamically, such as the number of local epochs and the batch size.
    \item \textbf{Enhance Privacy:} While FL can provide some privacy concerning the raw data of each user, the model updates can be reverse-engineered to reveal sensitive information about the users. Several techniques such as Differential privacy can be used to enhance the privacy-preservation in FL in vehicular environments.

\end{itemize}
}
\section{Conclusion}
In this paper, we have investigated the problem of clustered FL in vehicular networks. We aimed to fill the gap between clustering in vehicular networks and clustering in FL by designing a mobility-aware learning process for clustered FL. 
In the proposed architecture, we consider the v2v communication as an asset to overcome the communication bottleneck of FL in vehicular networks. Accordingly, in each communication round, a subset of vehicles are selected to act as cluster-heads, and the remainder of vehicles are matched the them. The selection favors vehicles with diverse datasets and good wireless communication channels with the gNodeB.  Furthermore, clustering based on the similarity of the updates is introduced to subdue the slow convergence of single joint FL model in non-i.i.d settings, especially in the presence of concept-shift. This step leads to the creation of new models which are sent to the non-participants and newly joint vehicles, who will evaluate them and score their preferences of these models. The resulting preference values are used to match each vehicle to their preferred model (cluster-head). 
Both the cluster-head selection and cluster matching are formulated as optimization problems with learning goals and mobility constraints.
We have proposed a greedy algorithm for the selection and RB allocation of cluster-heads, and a maximum weighted bipartite matching algorithm for the cluster formation. Simulations show the efficacy of using V2V communication to accelerate the learning as well as the importance of clustering based on updates to control concept shift. In the future, we aim to make the proposed approach resilient to outliers and malicious attacks such as false data injection.

%% file: Manuscript.bbl
\begin{thebibliography}{10}
\providecommand{\url}[1]{#1}
\csname url@samestyle\endcsname
\providecommand{\newblock}{\relax}
\providecommand{\bibinfo}[2]{#2}
\providecommand{\BIBentrySTDinterwordspacing}{\spaceskip=0pt\relax}
\providecommand{\BIBentryALTinterwordstretchfactor}{4}
\providecommand{\BIBentryALTinterwordspacing}{\spaceskip=\fontdimen2\font plus
\BIBentryALTinterwordstretchfactor\fontdimen3\font minus
  \fontdimen4\font\relax}
\providecommand{\BIBforeignlanguage}[2]{{%
\expandafter\ifx\csname l@#1\endcsname\relax
\typeout{** WARNING: IEEEtran.bst: No hyphenation pattern has been}%
\typeout{** loaded for the language `#1'. Using the pattern for}%
\typeout{** the default language instead.}%
\else
\language=\csname l@#1\endcsname
\fi
#2}}
\providecommand{\BIBdecl}{\relax}
\BIBdecl

\bibitem{yurtsever_survey_2020}
E.~Yurtsever, J.~Lambert, A.~Carballo, and K.~Takeda, ``A {Survey} of
  {Autonomous} {Driving}: {Common} {Practices} and {Emerging} {Technologies},''
  \emph{IEEE Access}, vol.~8, pp. 58\,443--58\,469, 2020, iEEE Access.

\bibitem{grigorescu_survey_2020}
\BIBentryALTinterwordspacing
S.~Grigorescu, B.~Trasnea, T.~Cocias, and G.~Macesanu,
  ``\BIBforeignlanguage{en}{A survey of deep learning techniques for autonomous
  driving},'' \emph{\BIBforeignlanguage{en}{Journal of Field Robotics}},
  vol.~37, no.~3, pp. 362--386, 2020, \_eprint:
  https://onlinelibrary.wiley.com/doi/pdf/10.1002/rob.21918. [Online].
  Available: \url{https://onlinelibrary.wiley.com/doi/abs/10.1002/rob.21918}
\BIBentrySTDinterwordspacing

\bibitem{lu_blockchain_2020}
Y.~Lu, X.~Huang, K.~Zhang, S.~Maharjan, and Y.~Zhang, ``Blockchain {Empowered}
  {Asynchronous} {Federated} {Learning} for {Secure} {Data} {Sharing} in
  {Internet} of {Vehicles},'' \emph{IEEE Transactions on Vehicular Technology},
  vol.~69, no.~4, pp. 4298--4311, Apr. 2020.

\bibitem{szabo_smartphone_2019}
L.~Szabó, L.~Lindenmaier, and V.~Tihanyi, ``Smartphone {Based} {HD} {Map}
  {Building} for {Autonomous} {Vehicles},'' in \emph{2019 {IEEE} 17th {World}
  {Symposium} on {Applied} {Machine} {Intelligence} and {Informatics}
  ({SAMI})}, Jan. 2019, pp. 365--370.

\bibitem{kim_cooperative_2016}
S.-W. Kim and W.~Liu, ``Cooperative {Autonomous} {Driving}: {A} {Mirror}
  {Neuron} {Inspired} {Intention} {Awareness} and {Cooperative} {Perception}
  {Approach},'' \emph{IEEE Intelligent Transportation Systems Magazine},
  vol.~8, no.~3, pp. 23--32, 2016.

\bibitem{filali_multi-access_2020}
A.~Filali, A.~Abouaomar, S.~Cherkaoui, A.~Kobbane, and M.~Guizani,
  ``Multi-{Access} {Edge} {Computing}: {A} {Survey},'' \emph{IEEE Access},
  vol.~8, pp. 197\,017--197\,046, 2020.

\bibitem{amine_mec}
A.~Abouaomar, S.~Cherkaoui, Z.~Mlika, and A.~Kobbane, ``Service function
  chaining in mec: A mean-field game and reinforcement learning approach,''
  2021.

\bibitem{r8}
\BIBentryALTinterwordspacing
J.~Konečný, H.~B. McMahan, D.~Ramage, and P.~Richtárik, ``Federated
  {Optimization}: {Distributed} {Machine} {Learning} for {On}-{Device}
  {Intelligence},'' \emph{arXiv:1610.02527 [cs]}, Oct. 2016. [Online].
  Available: \url{http://arxiv.org/abs/1610.02527}
\BIBentrySTDinterwordspacing

\bibitem{sattler_clustered_2020}
F.~Sattler, K.-R. Müller, and W.~Samek, ``Clustered {Federated} {Learning}:
  {Model}-{Agnostic} {Distributed} {Multitask} {Optimization} {Under} {Privacy}
  {Constraints},'' \emph{IEEE Transactions on Neural Networks and Learning
  Systems}, pp. 1--13, 2020.

\bibitem{briggs_federated_2020}
C.~Briggs, Z.~Fan, and P.~Andras, ``Federated learning with hierarchical
  clustering of local updates to improve training on non-{IID} data,'' Jul.
  2020, pp. 1--9, iSSN: 2161-4407.

\bibitem{meysam3}
M.~Azizian \emph{et~al.}, ``Dcev: A distributed cluster formation for vanet
  based on end-to-end realtive mobility,'' in \emph{2016 International Wireless
  Communications and Mobile Computing Conference (IWCMC)}, 2016, pp. 287--291.

\bibitem{yang_edge_2021}
B.~Yang, X.~Cao, K.~Xiong, C.~Yuen, Y.~L. Guan, S.~Leng, L.~Qian, and Z.~Han,
  ``Edge {Intelligence} for {Autonomous} {Driving} in {6G} {Wireless} {System}:
  {Design} {Challenges} and {Solutions},'' \emph{IEEE Wireless Communications},
  vol.~28, no.~2, pp. 40--47, Apr. 2021, iEEE Wireless Communications.

\bibitem{zeng_federated_2021}
\BIBentryALTinterwordspacing
T.~Zeng, O.~Semiari, M.~Chen, W.~Saad, and M.~Bennis,
  ``\BIBforeignlanguage{en}{Federated {Learning} on the {Road}: {Autonomous}
  {Controller} {Design} for {Connected} and {Autonomous} {Vehicles}},''
  \emph{\BIBforeignlanguage{en}{arXiv:2102.03401 [cs, eess]}}, Feb. 2021,
  arXiv: 2102.03401. [Online]. Available: \url{http://arxiv.org/abs/2102.03401}
\BIBentrySTDinterwordspacing

\bibitem{yu_mobility-aware_2020}
Z.~Yu, J.~Hu, G.~Min, Z.~Zhao, W.~Miao, and M.~S. Hossain, ``Mobility-{Aware}
  {Proactive} {Edge} {Caching} for {Connected} {Vehicles} {Using} {Federated}
  {Learning},'' \emph{IEEE Transactions on Intelligent Transportation Systems},
  pp. 1--11, 2020.

\bibitem{gps_fl}
Y.~Zhu, S.~Zhang, Y.~Liu, D.~Niyato, and J.~J. Yu, ``Robust federated learning
  approach for travel mode identification from non-iid gps trajectories,'' in
  \emph{2020 IEEE 26th International Conference on Parallel and Distributed
  Systems (ICPADS)}, 2020, pp. 585--592.

\bibitem{lim_federated_2020}
W.~Y.~B. Lim, N.~C. Luong, D.~T. Hoang, Y.~Jiao, Y.-C. Liang, Q.~Yang,
  D.~Niyato, and C.~Miao, ``Federated {Learning} in {Mobile} {Edge} {Networks}:
  {A} {Comprehensive} {Survey},'' \emph{IEEE Communications Surveys Tutorials},
  vol.~22, no.~3, pp. 2031--2063, 2020, iEEE Communications Surveys Tutorials.

\bibitem{aledhari_federated_2020}
M.~Aledhari, R.~Razzak, R.~M. Parizi, and F.~Saeed, ``Federated {Learning}: {A}
  {Survey} on {Enabling} {Technologies}, {Protocols}, and {Applications},''
  \emph{IEEE Access}, vol.~8, pp. 140\,699--140\,725, 2020, iEEE Access.

\bibitem{imteaj_survey_2021}
A.~Imteaj, U.~Thakker, S.~Wang, J.~Li, and M.~H. Amini, ``A {Survey} on
  {Federated} {Learning} for {Resource}-{Constrained} {IoT} {Devices},''
  \emph{IEEE Internet of Things Journal}, pp. 1--1, 2021, iEEE Internet of
  Things Journal.

\bibitem{elbir_federated_2020}
\BIBentryALTinterwordspacing
A.~M. Elbir, B.~Soner, and S.~Coleri, ``\BIBforeignlanguage{en}{Federated
  {Learning} in {Vehicular} {Networks}},''
  \emph{\BIBforeignlanguage{en}{arXiv:2006.01412 [cs, eess, math]}}, Sep. 2020,
  arXiv: 2006.01412. [Online]. Available: \url{http://arxiv.org/abs/2006.01412}
\BIBentrySTDinterwordspacing

\bibitem{meysam4}
M.~Azizian \emph{et~al.}, ``Vehicle software updates distribution with sdn and
  cloud computing,'' \emph{IEEE Communications Magazine}, vol.~55, no.~8, pp.
  74--79, 2017.

\bibitem{meysam5}
------, ``An optimized flow allocation in vehicular cloud,'' \emph{IEEE
  Access}, vol.~4, pp. 6766--6779, 2016.

\bibitem{jabri_vehicular_2019}
\BIBentryALTinterwordspacing
I.~Jabri, T.~Mekki, A.~Rachedi, and M.~Ben~Jemaa,
  ``\BIBforeignlanguage{en}{Vehicular fog gateways selection on the internet of
  vehicles: {A} fuzzy logic with ant colony optimization based approach},''
  \emph{\BIBforeignlanguage{en}{Ad Hoc Networks}}, vol.~91, p. 101879, Aug.
  2019. [Online]. Available:
  \url{https://www.sciencedirect.com/science/article/pii/S1570870518308096}
\BIBentrySTDinterwordspacing

\bibitem{al_ridhawi_continuous_2018}
\BIBentryALTinterwordspacing
I.~Al~Ridhawi, M.~Aloqaily, B.~Kantarci, Y.~Jararweh, and H.~T. Mouftah,
  ``\BIBforeignlanguage{en}{A continuous diversified vehicular cloud service
  availability framework for smart cities},''
  \emph{\BIBforeignlanguage{en}{Computer Networks}}, vol. 145, pp. 207--218,
  Nov. 2018. [Online]. Available:
  \url{https://www.sciencedirect.com/science/article/pii/S1389128618308430}
\BIBentrySTDinterwordspacing

\bibitem{tal_towards_2014}
I.~Tal and G.-M. Muntean, ``\BIBforeignlanguage{en}{Towards {Smarter} {Cities}
  and {Roads}: {A} {Survey} of {Clustering} {Algorithms} in {VANETs}},'' 2014,
  iSBN: 9781466659780 Pages: 16-50 Publisher: IGI Global.

\bibitem{cooper_comparative_2017}
C.~Cooper, D.~Franklin, M.~Ros, F.~Safaei, and M.~Abolhasan, ``A {Comparative}
  {Survey} of {VANET} {Clustering} {Techniques},'' \emph{IEEE Communications
  Surveys Tutorials}, vol.~19, no.~1, pp. 657--681, 2017, iEEE Communications
  Surveys Tutorials.

\bibitem{meysam1}
M.~Azizian \emph{et~al.}, ``A distributed d-hop cluster formation for vanet,''
  in \emph{2016 IEEE Wireless Communications and Networking Conference}, 2016,
  pp. 1--6.

\bibitem{meysam2}
------, ``A distributed cluster based transmission scheduling in vanet,'' in
  \emph{2016 IEEE International Conference on Communications (ICC)}, 2016, pp.
  1--6.

\bibitem{singh_nwca_2016}
\BIBentryALTinterwordspacing
D.~Singh, Ranvijay, and R.~S. Yadav, ``{NWCA}: {A} {New} {Weighted}
  {Clustering} {Algorithm} to form {Stable} {Cluster} in {VANET},'' in
  \emph{Proceedings of the {Second} {International} {Conference} on
  {Information} and {Communication} {Technology} for {Competitive}
  {Strategies}}, ser. {ICTCS} '16.\hskip 1em plus 0.5em minus 0.4em\relax New
  York, NY, USA: Association for Computing Machinery, Mar. 2016, pp. 1--6.
  [Online]. Available: \url{https://doi.org/10.1145/2905055.2905226}
\BIBentrySTDinterwordspacing

\bibitem{daeinabi_vwca_2011}
\BIBentryALTinterwordspacing
A.~Daeinabi, A.~G. Pour~Rahbar, and A.~Khademzadeh,
  ``\BIBforeignlanguage{en}{{VWCA}: {An} efficient clustering algorithm in
  vehicular ad hoc networks},'' \emph{\BIBforeignlanguage{en}{Journal of
  Network and Computer Applications}}, vol.~34, no.~1, pp. 207--222, Jan. 2011.
  [Online]. Available:
  \url{https://www.sciencedirect.com/science/article/pii/S1084804510001384}
\BIBentrySTDinterwordspacing

\bibitem{kim_dynamic_2020}
\BIBentryALTinterwordspacing
Y.~Kim, E.~A. Hakim, J.~Haraldson, H.~Eriksson, J.~M. B.~d. Silva~Jr., and
  C.~Fischione, ``\BIBforeignlanguage{en}{Dynamic {Clustering} in {Federated}
  {Learning}},'' \emph{\BIBforeignlanguage{en}{arXiv:2012.03788 [cs]}}, Dec.
  2020, arXiv: 2012.03788. [Online]. Available:
  \url{http://arxiv.org/abs/2012.03788}
\BIBentrySTDinterwordspacing

\bibitem{ghosh_efficient_2021}
\BIBentryALTinterwordspacing
A.~Ghosh, J.~Chung, D.~Yin, and K.~Ramchandran, ``\BIBforeignlanguage{en}{An
  {Efficient} {Framework} for {Clustered} {Federated} {Learning}},''
  \emph{\BIBforeignlanguage{en}{arXiv:2006.04088 [cs, stat]}}, Jun. 2021,
  arXiv: 2006.04088. [Online]. Available: \url{http://arxiv.org/abs/2006.04088}
\BIBentrySTDinterwordspacing

\bibitem{kairouz_advances_2021}
\BIBentryALTinterwordspacing
P.~Kairouz \emph{et~al.}, ``Advances and {Open} {Problems} in {Federated}
  {Learning},'' \emph{arXiv:1912.04977 [cs, stat]}, Mar. 2021, arXiv:
  1912.04977. [Online]. Available: \url{http://arxiv.org/abs/1912.04977}
\BIBentrySTDinterwordspacing

\bibitem{tak_federated_2021}
A.~Tak and S.~Cherkaoui, ``Federated {Edge} {Learning}: {Design} {Issues} and
  {Challenges},'' \emph{IEEE Network}, vol.~35, no.~2, pp. 252--258, Mar. 2021.

\bibitem{mansour_three_2020}
\BIBentryALTinterwordspacing
Y.~Mansour, M.~Mohri, J.~Ro, and A.~T. Suresh, ``Three {Approaches} for
  {Personalization} with {Applications} to {Federated} {Learning},''
  \emph{arXiv:2002.10619 [cs, stat]}, Jul. 2020, arXiv: 2002.10619. [Online].
  Available: \url{http://arxiv.org/abs/2002.10619}
\BIBentrySTDinterwordspacing

\bibitem{chen_zero_2021}
Z.~Chen, P.~Tian, W.~Liao, and W.~Yu, ``Zero {Knowledge} {Clustering} {Based}
  {Adversarial} {Mitigation} in {Heterogeneous} {Federated} {Learning},''
  \emph{IEEE Transactions on Network Science and Engineering}, vol.~8, no.~2,
  pp. 1070--1083, Apr. 2021, iEEE Transactions on Network Science and
  Engineering.

\bibitem{taik_data-quality_2021}
A.~Taïk, H.~Moudoud, and S.~Cherkaoui, ``Data-{Quality} {Based} {Scheduling}
  for {Federated} {Edge} {Learning},'' in \emph{2021 {IEEE} 46th {Conference}
  on {Local} {Computer} {Networks} ({LCN})}, Oct. 2021, pp. 17--23, iSSN:
  0742-1303.

\bibitem{liu_client-edge-cloud_2020}
L.~Liu, J.~Zhang, S.~Song, and K.~B. Letaief, ``Client-{Edge}-{Cloud}
  {Hierarchical} {Federated} {Learning},'' in \emph{{ICC} 2020 - 2020 {IEEE}
  {International} {Conference} on {Communications} ({ICC})}, Jun. 2020, pp.
  1--6, iSSN: 1938-1883.

\bibitem{chai_hierarchical_2021}
H.~Chai, S.~Leng, Y.~Chen, and K.~Zhang, ``A {Hierarchical}
  {Blockchain}-{Enabled} {Federated} {Learning} {Algorithm} for {Knowledge}
  {Sharing} in {Internet} of {Vehicles},'' \emph{IEEE Transactions on
  Intelligent Transportation Systems}, vol.~22, no.~7, pp. 3975--3986, Jul.
  2021, iEEE Transactions on Intelligent Transportation Systems.

\bibitem{taik_data-aware_2021}
A.~Taïk, Z.~Mlika, and S.~Cherkaoui, ``Data-{Aware} {Device} {Scheduling} for
  {Federated} {Edge} {Learning},'' \emph{IEEE Transactions on Cognitive
  Communications and Networking}, pp. 1--1, 2021.

\bibitem{ren_unified_2018}
M.~Ren, J.~Zhang, L.~Khoukhi, H.~Labiod, and V.~Vèque, ``A {Unified}
  {Framework} of {Clustering} {Approach} in {Vehicular} {Ad} {Hoc}
  {Networks},'' \emph{IEEE Transactions on Intelligent Transportation Systems},
  vol.~19, no.~5, pp. 1401--1414, May 2018, iEEE Transactions on Intelligent
  Transportation Systems.

\bibitem{li_robust_2012}
W.~Li, A.~Tizghadam, and A.~Leon-Garcia, ``Robust clustering for connected
  vehicles using local network criticality,'' in \emph{2012 {IEEE}
  {International} {Conference} on {Communications} ({ICC})}, Jun. 2012, pp.
  7157--7161, iSSN: 1938-1883.

\bibitem{ren_mobility-based_2017}
\BIBentryALTinterwordspacing
M.~Ren, L.~Khoukhi, H.~Labiod, J.~Zhang, and V.~Vèque, ``A mobility-based
  scheme for dynamic clustering in vehicular ad-hoc networks ({VANETs}),''
  vol.~9, pp. 233--241, 2017. [Online]. Available:
  \url{https://www.sciencedirect.com/science/article/pii/S2214209616300699}
\BIBentrySTDinterwordspacing

\bibitem{pisinger_where_2005}
\BIBentryALTinterwordspacing
D.~Pisinger, ``\BIBforeignlanguage{en}{Where are the hard knapsack problems?}''
  \emph{\BIBforeignlanguage{en}{Computers \& Operations Research}}, vol.~32,
  no.~9, pp. 2271--2284, Sep. 2005. [Online]. Available:
  \url{https://www.sciencedirect.com/science/article/pii/S030505480400036X}
\BIBentrySTDinterwordspacing

\bibitem{cormen_introduction_2009}
T.~H. Cormen, C.~E. Leiserson, R.~L. Rivest, and C.~Stein,
  \emph{\BIBforeignlanguage{en}{Introduction to {Algorithms}}}.\hskip 1em plus
  0.5em minus 0.4em\relax MIT Press, Jul. 2009.

\bibitem{chen_group-aware_2016}
\BIBentryALTinterwordspacing
C.~Chen, S.~Chester, V.~Srinivasan, K.~Wu, and A.~Thomo, ``Group-{Aware}
  {Weighted} {Bipartite} {B}-{Matching},'' in \emph{Proceedings of the 25th
  {ACM} {International} on {Conference} on {Information} and {Knowledge}
  {Management}}, ser. {CIKM} '16.\hskip 1em plus 0.5em minus 0.4em\relax New
  York, NY, USA: Association for Computing Machinery, Oct. 2016, pp. 459--468.
  [Online]. Available: \url{https://doi.org/10.1145/2983323.2983770}
\BIBentrySTDinterwordspacing

\bibitem{matching2017}
\BIBentryALTinterwordspacing
F.~Ahmed, J.~P. Dickerson, and M.~Fuge, ``Diverse weighted bipartite
  b-matching,'' \emph{Proceedings of the Twenty-Sixth International Joint
  Conference on Artificial Intelligence}, Aug 2017. [Online]. Available:
  \url{http://dx.doi.org/10.24963/ijcai.2017/6}
\BIBentrySTDinterwordspacing

\bibitem{noauthor_optimization_nodate}
\BIBentryALTinterwordspacing
``Optimization with {PuLP} — {PuLP} 2.5.0 documentation.'' [Online].
  Available: \url{https://coin-or.github.io/pulp/}
\BIBentrySTDinterwordspacing

\bibitem{noauthor_pytorch_nodate}
\BIBentryALTinterwordspacing
``\BIBforeignlanguage{en}{{PyTorch}}.'' [Online]. Available:
  \url{https://www.pytorch.org}
\BIBentrySTDinterwordspacing

\bibitem{mnist}
Y.~Lecun, L.~Bottou, Y.~Bengio, and P.~Haffner, ``Gradient-based learning
  applied to document recognition,'' \emph{Proceedings of the IEEE}, vol.~86,
  no.~11, pp. 2278--2324, Nov. 1998.

\bibitem{xiao_fashion-mnist_2017}
\BIBentryALTinterwordspacing
H.~Xiao, K.~Rasul, and R.~Vollgraf, ``Fashion-{MNIST}: a {Novel} {Image}
  {Dataset} for {Benchmarking} {Machine} {Learning} {Algorithms},''
  \emph{arXiv:1708.07747 [cs, stat]}, Sep. 2017, arXiv: 1708.07747. [Online].
  Available: \url{http://arxiv.org/abs/1708.07747}
\BIBentrySTDinterwordspacing

\bibitem{tolpegin_data_2020}
\BIBentryALTinterwordspacing
V.~Tolpegin, S.~Truex, M.~E. Gursoy, and L.~Liu, ``Data {Poisoning} {Attacks}
  {Against} {Federated} {Learning} {Systems},'' \emph{arXiv:2007.08432 [cs,
  stat]}, Aug. 2020, arXiv: 2007.08432. [Online]. Available:
  \url{http://arxiv.org/abs/2007.08432}
\BIBentrySTDinterwordspacing

\bibitem{liang_resource_2017}
L.~Liang, G.~Y. Li, and W.~Xu, ``Resource {Allocation} for {D2D}-{Enabled}
  {Vehicular} {Communications},'' \emph{IEEE Transactions on Communications},
  vol.~65, no.~7, pp. 3186--3197, Jul. 2017, note: IEEE Transactions on
  Communications.

\bibitem{hajar_blockchain1}
H.~Moudoud, S.~Cherkaoui, and L.~Khoukhi, ``Towards a scalable and trustworthy
  blockchain: Iot use case,'' in \emph{ICC 2021 - IEEE International Conference
  on Communications}, 2021, pp. 1--6.

\end{thebibliography}
